\documentclass[english,aps,prd,reprint,showpacs,floatfix,unsortedaddress]{revtex4-1}

\usepackage[T1]{fontenc}
\usepackage[latin1]{inputenc}
\usepackage[english]{babel}
\usepackage{amsmath}
\usepackage{graphicx}
\usepackage{amssymb}

\makeatother

\makeatletter

\begin{document}

\title{Thermodynamics of quantum photon spheres}

\author{M. C. Baldiotti}
\email{baldiotti@uel.br}
\affiliation{Departamento de F\'{\i}sica, Universidade Estadual de Londrina, 86051-990, Londrina, Paran\'{a}, Brazil}

\author{Walace S. Elias}
\email{walace.elias@usp.br}
\affiliation{Instituto de F\'{\i}sica, Universidade de S\~{a}o Paulo,
 C.P. 66318, 05315-970, S\~{a}o Paulo-SP, Brazil}

\author{C. Molina}
\email{cmolina@usp.br}
\affiliation{Escola de Artes, Ci\^{e}ncias e Humanidades, Universidade de S\~{a}o Paulo, Av. Arlindo Bettio 1000, 03828-000, S\~{a}o Paulo-SP, Brazil}

\author{Thiago S. Pereira}
\email{tspereira@uel.br}
\affiliation{Departamento de F\'{\i}sica, Universidade Estadual de Londrina, 86051-990, Londrina, Paran\'{a}, Brazil}

\begin{abstract}

Photon spheres, surfaces where massless particles are confined in closed orbits, are expected to be common astrophysical structures surrounding ultracompact objects. In this paper a semiclassical treatment of a photon sphere is proposed. We consider the quantum Maxwell field and derive its energy spectra. A thermodynamic approach for the quantum photon sphere is developed and explored. Within this treatment, an expression for the spectral energy density of the emitted radiation is presented. Our results suggest that photon spheres, when thermalized with their environment, have nonusual thermodynamic properties, which could lead to distinct observational signatures. 

\end{abstract}


%

\maketitle

\section{Introduction}

General relativity predicts the existence of regions where light is confined in closed orbits. These structures, the so-called photon spheres or light spheres, are expected to be common astrophysical objects surrounding ultracompact bodies \cite{0264-9381-2-2-013,Nemiroff:1993zz,Nemiroff:1994,Narayan:2005ie,PhysRevD.90.044069}. Although black holes are natural candidates to create light lines and light surfaces, other bodies could support such objects. Initially considered as a particular feature of the Schwarzschild spacetime, the photon sphere concept was generalized and found in a broad class of static and spherically symmetric geometries \cite{Claudel:2000yi,Foertsch:2003ze}. Given an approximate spherical symmetry, staticity, and reasonable energy conditions, photon spheres should be present, even considering extensions of Einstein's relativity \cite{Claudel:2000yi,Foertsch:2003ze}.

More recently, it is observed a renewed interest in the physics of photon spheres. For instance, the problem of the characterization of the photon spheres in several geometries was treated, for example, in \cite{Claudel:2000yi,Foertsch:2003ze,Horvat:2013plm}. The connection between photon sphere parameters and quantities associated to the perturbative dynamics around black holes has been recently explored, for example, in \cite{PhysRevD.79.064016,PhysRevD.81.104039,PhysRevD.90.044069}. From the observational point of view, light sphere and light ring phenomenology is also an issue. For instance, light ring astrophysical signatures are explored in \cite{Moscibrodzka:2009gw,Johannsen:2010ru}.

The physical framework we consider is commented in the following. We assume that a spherically symmetric and static distribution of matter generates a photon sphere, with the photons propagating in vacuum or in optically transparent media. We also assume some exchange of photons of the photon sphere with the surrounding environment, in such a way that the photon sphere is in thermal equilibrium with the environment. The photons are considered as metastable entities, approximately free bosonic particles with a finite average lifetime in the photon sphere.

We propose a semiclassical treatment for quantum photons in light spheres, with the quantized electromagnetic field in a classical photon sphere background. By considering Maxwell's electrodynamics in usual spherical coordinates and in a suitable gauge, energy spectra for the quantum physical modes are derived. Our approach suggests that light spheres populated by photons in thermal equilibrium with their environment have distinct thermodynamic properties, which could lead to observable signatures. 

The structure of this paper is presented in the following. In Sec.~\ref{sec_lightspheres} we review the notation and comment on some key characteristics of the photon spheres, emphasizing the classical energy spectrum of this system. A quantum treatment for the electromagnetic field in the photon sphere is introduced in Sec.~\ref{sec_spectrum}. Quantum energy spectra are obtained for the two physical polarizations of the field, and the connections with the classical limit are discussed. In Sec.~\ref{sec_thermodynamics} the photon sphere is presented as a thermal bosonic system, and its thermodynamics is characterized. Some final remarks are made in Sec.~\ref{sec_final}. We use signature $(-,+,+,+)$ and natural units with $G=\hbar=c=k_{B}=1$ throughout this paper.

\section{Photon spheres in spherically symmetric spacetimes}
\label{sec_lightspheres}

In the present work, we are interested in static and spherically symmetric spacetimes. These geometries are equipped with four Killing vector fields $\mathcal{K}_{t}$, $\mathcal{K}_{x}$, $\mathcal{K}_{y}$, and $\mathcal{K}_{z}$ satisfying
\begin{equation}
\left[ \mathcal{K}_{x},\mathcal{K}_{y}\right] = \mathcal{K}_{z}
\,\,,\,\,
\left[
\mathcal{K}_{y},\mathcal{K}_{z}\right] = \mathcal{K}_{x}
\,\,,\,\,
\left[
\mathcal{K}_{z},\mathcal{K}_{x}\right] = \mathcal{K}_{y}
\,\,,
\label{so3_killings}
\end{equation}
and
\begin{equation}
\left[ \mathcal{K}_{t},\mathcal{K}_{x}\right]  
= \left[ \mathcal{K}_{t},\mathcal{K}_{y}\right] 
= \left[ \mathcal{K}_{t},\mathcal{K}_{z}\right]
= 0 \,\, .
\label{comuta_t_rotacao}
\end{equation}
We are assuming the existence of a ``static region'' in the spacetime, where $\mathcal{K}_{x}$, $\mathcal{K}_{y}$, $\mathcal{K}_{z}$ are spacelike and $\mathcal{K}_{t}$ is timelike. For these spacetimes, a coordinate system $(t,r,\theta,\phi)$ may be defined in the static region, such that $\mathcal{K}_{t} = \partial/\partial t$, $r$ is the ``areal radius'' and $(\theta$,$\phi)$ are the usual angular coordinates that cover $S^{2}$ surfaces, invariant under the action of the diffeomorphisms associated to $\mathcal{K}_{x}$, $\mathcal{K}_{y}$, and $\mathcal{K}_{z}$. In terms of this coordinate system, the line element is written as
\begin{equation}
ds^{2} = g_{tt}(r) \, dt^{2} + g_{rr}(r) \, dr^{2} 
+ r^{2}\left( d\theta^{2} + \sin^{2} \theta\, d\phi^{2} \right) \,\, ,
\label{metric_rt}
\end{equation}
with 
$-g_{tt}(r)>0$ and $g_{rr}(r)>0$
in the static region. 

We will also assume that the spacetime is asymptotically flat and that the compact object modeled by this geometry has no electric charge. In this case, the metric functions behave as 
$-g_{tt}(r) \rightarrow 1$ and $g_{rr}^{-1}(r) \rightarrow 1$ 
in the limit $r\rightarrow\infty$. 
Considering the spacetime described by Eq.~\eqref{metric_rt}, the photon sphere is a two dimensional surface generated by null geodesics that describe circular orbits. The photon sphere radius is denoted in the present work as $R$. If the geometry models a black hole, the photon sphere will be located outside the event horizon, always in the  static region. The spacetimes selected by the conditions imposed include not only the Schwarzschild geometry but also many other cases of interest.  

Given affine parametrized null geodesics $x^{\mu}(\lambda)$, four constants of motion can be constructed: $E$, $L_{x}$, $L_{y}$, and $L_{z}$, associated to the Killing fields $\mathcal{K}_{t}$, $\mathcal{K}_{x}$, $\mathcal{K}_{y}$, and $\mathcal{K}_{z}$, respectively. If the geometry is asymptotically flat, these constants can be interpreted as energy and angular momentum components associated to the geodesic, as seen by a static observer far from the compact object. For geodesics in the photon sphere,  $E$, $L_{x}$, $L_{y}$, and $L_{z}$ obey the classical constraint \cite{Chandra1998mathematical}:
\begin{equation}
E^{2} = - g_{tt}(R) \, \frac{L^{2}}{R^{2}} \,\, ,
\label{classical_constraint}
\end{equation}
with $L^{2} = L_{x}^{2} + L_{y}^{2} + L_{z}^{2}$. The constant $L^{2}$ can be interpreted as the (classical) squared total angular momentum of the photon in the geodesic.

For any given geodesic in the photon sphere, it is always possible to choose a coordinate system such that this null orbit is located in the equatorial plane ($\theta = \pi/2$). In this case, the equation of motion which describes the geodesic is \cite{Chandra1998mathematical}
\begin{equation}
\left[ \frac{dr(\lambda)}{d\lambda} \right]^{2} = V_{\textrm{eff}}(r(\lambda))   \,\, ,
\label{eq_motion}
\end{equation}
where the effective potential $V_{eff}$ is given by
\begin{equation}
V_{\textrm{eff}}(r) =  \frac{1}{g_{rr}(r)} \left[ + \frac{E^{2}}{-g_{tt}(r)}  - \frac{L^{2}}{r^{2}} \right] \,\, .
\label{Veff}
\end{equation}
The photon sphere radius $R$ is such that $V_{\textrm{eff}}(R)=0$ and $dV_{\textrm{eff}}(R)/dr=0$. A basic fact about photon spheres is that the classical trajectories that form this structure are usually, but not necessarily, unstable. Spacetimes where the photon sphere orbits are stable are presented, for example, in \cite{Karlovini:2000xd,Hasse:2001by}. 
For unstable photon orbits, it is possible to estimate an average lifetime $\tau$ for the null circular geodesics \cite{PhysRevD.79.064016},
\begin{equation}
\tau = \left[
\frac{- g_{tt}(R) R^{2}}{2 L^{2}} \left. \frac{d^{2} V_{\textrm{eff}}(r)}{dr^{2}} \right|_{r=R} 
\right]^{-1/2}
\,\,.
\label{lifetime}
\end{equation}

However, it should be pointed out that the average lifetime $\tau$ must be used cautiously. It is possible to show that, in spacetimes of interest, the time averaged Lyapunov exponent vanishes and the orbit behaves as if there were no instability, even though it clearly has an unstable region in phase space (this is a general feature of a larger class of dynamical systems involving the so-called homoclinic orbits \cite{Belbruno:2011nn,Cornish:2003ig}). 
As a result, photons in the light sphere may have long lifetimes, even when they are characterized by unstable effective potentials. 

At this point, although we are not restricted to this specific scenario, it is illustrative to particularize our discussion considering the photon sphere prototype, present in the Schwarzschild spacetime. For this geometry, 
$-g_{tt}(r) = g_{rr}^{-1}(r) = 1 - 2M/r$. 
In this case $R=3M$, and therefore a photon sphere can be supported by a spherical compact body of mass $M$, with a radius smaller than $3M$. The Schwarzschild black hole is one possibility for such an object. An estimate for the lifetime of a photon in this photon sphere is $\tau=3\sqrt{3}M$. 
The relevant point about this calculated value for $\tau$ is the fact that it is finite and proportional to the compact body mass. But again, it should be used cautiously. Null geodesics in Schwarzschild spacetime are one example where the averaged Lyapunov exponent vanishes \cite{Belbruno:2011nn,Cornish:2003ig}. More concretely, null geodesics spiral around $r=R$ if the classical constraint in Eq.~\eqref{classical_constraint} is approximately obeyed, as can be seen by considering the solutions presented in Eqs.~(231) and (238) in Chap.~3 of \cite{Chandra1998mathematical}, for instance. In this case, the null-mass particles may have quasicircular trajectories, circling many times around the photon sphere, arbitrarily close the surface $r=R$ \cite{Chandra1998mathematical}.
The preceding example shows that photons can be expected to be in the light spheres for quite long times in relevant astrophysical scenarios, such as large black holes in vacuum.

\section{Energy spectra of photon spheres}
\label{sec_spectrum}

The picture of quantum photons around a noncharged compact object will be made more precise considering astrophysical situations where the photons in the light sphere have a long lifetime, and therefore are essentially confined to the surface $r=R$. The photon sphere is in the outside region of the black hole (if one is present) and the radiation emitted by the photon sphere is detected by a distant observer.  As a consequence, issues associated to the field behavior at the event horizon \cite{PhysRevD.33.1590} are not a problem here.  We will effectively quantize the electromagnetic field in the three dimensional manifold $S^{2} \times \mathbb{R}$. 

The typical strategy in the quantization of the electromagnetic field, based on plane wave expansions, is not convenient to us due to the geometry of the photon sphere. For our purposes, a spherical representation of the field is better suited. However, given the gauge arbitrariness, it is not obvious that in this representation the electromagnetic field can be decomposed in two independent modes (polarizations). This issue was already considered in \cite{Ruffini:1972pw,Chandra1998mathematical} in classical contexts involving curved spacetimes. The quantum field theory treatment was developed, for example, in \cite{0264-9381-4-3-029,PhysRevD.57.1108,PhysRevD.57.1108,PhysRevD.63.124008,PhysRevD.80.029906}. In this paper we will adapt the treatment developed in \cite{0264-9381-4-3-029,PhysRevD.63.124008}, considering the electromagnetic quanta in the photon sphere. 

We proceed to the quantum treatment with the introduction of the Maxwell field,  minimally coupled to the geometry. From the classical electromagnetic tensor $F_{\mu\nu}$, the potential $A_{\mu}$ is defined as 
\begin{equation}
F_{\mu\nu} = \nabla_{\mu} A_{\nu} - \nabla_{\nu} A_{\mu} \,\, .
\end{equation}
Following \cite{PhysRevD.63.124008}, we adopt a modified Feynman gauge, with the Lagrangian density written as
\begin{equation}
\mathcal{L}_{F} = \sqrt{-g} \left[ -\frac{1}{4} F_{\mu\nu} F^{\mu\nu} 
- \frac{1}{2} G^{2}
\right] \,\, ,
\label{Lagrangian}
\end{equation}
where $g=\det \, \left[ g^{\mu\nu} \right]$ and
\begin{equation}
G = \nabla^{\mu} A_{\mu} + K^{\mu} A_{\mu} \,\, ,
\label{G_function}
\end{equation}
\begin{equation}
K^{\mu} = (0,g_{tt}'(r)/[g_{tt}(r)\,g_{rr}(r)],0,0 ) \,\, . 
\label{K_field}
\end{equation}
The gauge condition then reads $G=0$.

For our purposes, only the physical modes (according to \cite{PhysRevD.63.124008}) are relevant. An important result is that, considering the modified Feynman gauge defined by \linebreak Eqs.~\eqref{Lagrangian}--\eqref{K_field}, the electromagnetic potential has a scalar and a vector physical modes %
\footnote{In \cite{PhysRevD.63.124008}, scalar and vector modes are called ``physical mode I'' and ``physical mode II,'' respectively.}
. These ``polarizations'' will be observables of the theory. 
In the photon sphere, the scalar mode potential $A_{\mu}^{sc}$  has only one non-null component,
\begin{equation}
A_{\mu}^{sc} = (0,A_{r}^{sc},0,0) \,\, .
\end{equation}
On the other hand, the vector mode potential $A_{\mu}^{vec}$ can be written as
\begin{equation}
A_{\mu}^{vec} = (0,0,A_{\theta}^{vec},A_{\phi}^{vec}) \,\, .
\end{equation}
In the following, we will proceed to the quantization of the electromagnetic field in $r=R$.

\subsection{Scalar and vector physical modes}

For the quantization of the potentials $\hat{A}_{r}^{sc}$ and $\hat{A}_{i}^{vec}$, corresponding respectively to the scalar and vector sectors in the photon sphere, we construct the one-particle Hilbert spaces 
\footnote{For the one-particle sector we mean the quantum mechanics with the finite product of Hilbert spaces instead of the Fock version in the completion of these spaces, i.e., before the second quantization process \cite{wald1984general}.}
$\mathcal{H}_{1}^{sc}$ and $\mathcal{H}_{1}^{vec}$ associated to the (scalar and vector mode) photons. We take as $\mathcal{H}_{1}^{sc}$ the set of functions $\mathcal{H}_{1}^{sc}:\,S^{2}\rightarrow\mathbb{C}$ that have, as a dense subset, the collection of functions $f$ that can be expanded as
\begin{equation}
f = \sum_{\ell m}\,f_{\ell m}(t)\,Y_{\ell m}(\theta,\phi) \,\, ,
\label{forma_Phi}
\end{equation}
where $Y_{\ell m}$ are the spherical harmonics.

In a similar way, we define the one-particle Hilbert space $\mathcal{H}_{1}^{vec}$, associated to the (vector mode) photons, as the set of functions
$\mathcal{H}_{1}^{vec}:\text{vec}\left(  S^{2}\right)  \rightarrow\mathbb{C}$
that have, as a dense subset, the collection of vectors $\tilde{f}_{i}$ that can be expanded in vector spherical harmonics,
\begin{equation}
\tilde{f}_{i} = \sum_{\ell m}\,\tilde{f}_{\ell m}(t)\,Y_{i}^{(\ell m)} (\theta,\phi) \,\,,
\end{equation}
with $\text{vec}\left(  S^{2} \right)$ denoting the $S^{2}$ vector bundle and $Y_{i}^{(\ell m)}$ the vector spherical harmonics \cite{0264-9381-4-3-029,PhysRevD.63.124008}.

With $\mathcal{H}_{1}^{p}$ defined, for $p = sc$ for the scalar mode and $p = vec$ for the vector mode, the one-particle Hamiltonian can be constructed, based on the Killing vector field $\mathcal{K}_{t}$, as
\begin{equation}
\hat{H}^{p} = \mathrm{i} \, \frac{\partial}{\partial t} \,\, ,
\,\, p \in \{ sc,vec \}  \,\, .
\label{H_explicit}
\end{equation}
That is, our notion of energy is being defined by static observers which follow integral curves of $\mathcal{K}_{t}$. The one-particle angular momentum operators $\hat{L}_{x}^{p}$, $\hat{L}_{y}^{p}$ and $\hat{L}_{z}^{p}$ are given by
\begin{align}
\hat{L}_{x}^{p}  & = \mathrm{i} \, \left[  -\sin\phi\,\frac{\partial}{\partial\theta} - \cot\theta\,\cos\phi\,\frac{\partial}{\partial\phi} \right] \,\, ,
\label{Lx_explicit}\\
\hat{L}_{y}^{p}  & = \mathrm{i} \, \left[  \cos\phi\,\frac{\partial}{\partial\theta} -\cot\theta\,\sin\phi\,\frac{\partial}{\partial\phi}\right] \,\, ,
\label{Ly_explicit}\\
\hat{L}_{z}^{p}  & = \mathrm{i} \, \frac{\partial}{\partial\phi} \,\, ,
\label{Lz_explicit}
\end{align}
where $p \in \{ sc,vec \}$ in Eqs.~\eqref{Lx_explicit}--\eqref{Lz_explicit}.

From the one-particle sector, we construct the Fock space associated with the photons in the photon sphere with the usual procedure. Taking into account Eq.~\eqref{H_explicit}, we observe that 
\begin{equation}
Y_{\ell m}(\theta,\phi) \, e^{-\mathrm{i} \epsilon t} 
\,\, , \,\, 
Y_{\ell m}^{\ast}(\theta,\phi) \, e^{+\mathrm{i} \epsilon t}
\,\, ,
\end{equation}
are positive and negative energy modes which span a dense subset of $\mathcal{H}_{1}^{sc}$. We define creation and annihilation operators, $(\hat{a}_{\epsilon\ell m}^{sc})^{\dagger}$ and $\hat{a}_{\epsilon\ell m}^{sc}$, such that the quantum version of the electromagnetic potential in the photon sphere, the operator $\hat{A}_{r}^{sc}$, can be expanded as
\begin{equation}
\hat{A}_{r}^{sc} = \sum_{j} \, \left[  \hat{a}_{(j)}^{sc} \, Y_{\ell m}(\theta,\phi)
\, e^{-\mathrm{i}\epsilon t} + \left(  \hat{a}_{(j)}^{sc} \right)^{\dagger} \,
Y_{\ell m}^{\ast}(\theta,\phi)\,e^{+\mathrm{i}\epsilon t}\right]
,
\label{hatA_expansion}
\end{equation}
with $(j) \equiv ( \epsilon \, \ell \, m )$. In addition, based on the definition of $\mathcal{H}_{1}^{vec}$, we note that
\begin{equation}
Y_{i}^{(\ell m)}(\theta,\phi) \, e^{-\mathrm{i}\epsilon t} \,\, , \,\,
(Y_{i}^{\ell m})^{*} (\theta,\phi) \, e^{+\mathrm{i}\epsilon t} \,\, ,
\end{equation}
are positive and negative energy modes which span a dense subset of 
$\mathcal{H}_{1}^{vec}$. Creation and annihilation operators 
$(\hat{a}_{\epsilon\ell m}^{vec})^{\dagger}$ 
and 
$\hat{a}_{\epsilon\ell m}^{vec}$ are introduced such that $\hat{A}_{i}^{vec}$, with $i\in \{\theta,\phi\}$, can be expanded as
%
%
\begin{eqnarray}
\hat{A}_{i}^{vec}  & = & \sum_{j} \, \Big[
\hat{a}_{(j)}^{vec}\,Y_{i}^{(\ell m)}(\theta,\phi)\,e^{-\mathrm{i}\epsilon t}
\nonumber\\
&& + \left(  \hat{a}_{(j)}^{vec}\right)  ^{\dagger}\,\left(  Y_{i}^{(\ell m)}\right)  ^{\ast}(\theta,\phi)\,e^{+\mathrm{i}\epsilon t} \Big] \,\, ,
\label{hatA_vec_expansion}
\end{eqnarray}
with $(j)\equiv(\epsilon\,\ell\,m)$, consonant to the notation in Eq.~\eqref{hatA_expansion}.

According to the usual conventions, we denote by $\hat{H}^{p}$, $\hat{L}_{x}^{p}$, $\hat{L}_{y}^{p}$, and $\hat{L}_{z}^{p}$ the Hamiltonian and (orbital) angular momentum operators acting in the Fock space. Distinction from their one-particle counterparts is made by context. With the Casimir operator $(\hat{L}^{p})^{2}$ given by 
\begin{equation}
(\hat{L}^{p})^{2} = (\hat{L}_{x}^{p})^{2} + (\hat{L}_{y}^{p})^{2} + (\hat{L}_{z}^{p})^{2} \,\, ,
\end{equation}
expressions for $\hat{H}^{p}$ and $(\hat{L}^{p})^{2}$ in terms of the creation and annihilation operators $\hat{a}_{(i)}^{p}$ and $(\hat{a}_{(i)}^{p})^{\dagger}$ are
\begin{eqnarray}
\hat{H}^{p}  & = & \sum_{i} \epsilon \, \left(  
\hat{a}_{(i)}^{p}\right)^{\dagger}
\hat{a}_{(i)}^{p} \,\, ,
\label{H_a_adagger} \\
\left( \hat{L}^{sc}\right)^{2}  & = & \sum_{i}\ell(\ell+1) \, 
\left( \hat{a}_{(i)}^{sc}\right)^{\dagger} 
\hat{a}_{(i)}^{sc} \,\, ,
\label{L2_a_adagger} \\
\left(  \hat{L}^{vec}\right)^{2}  & = & \sum_{i}\left[  \ell(\ell+1)- 1 \right]
\, \left(  \hat{a}_{(i)}^{vec}\right)^{\dagger} \hat{a}_{(i)}^{vec}
\,\, .
\label{L2_a_adagger_vec}
\end{eqnarray}
Quanta defined by $( \hat{a}_{(i)}^{p} )^{\dagger}$ and $\hat{a}_{(i)}^{p}$ have well-defined energy $\epsilon$ and squared angular momentum, the later quantity with magnitude $\ell(\ell+1)$ for the scalar and $[\ell(\ell+1)-1]$ for the vector modes.

We set as the quantum constraint for the scalar and vector sectors of the multiparticle photon sphere
\begin{equation}
[ \hat{H}^{p}, [\hat{H}^{p} , \hat{A}_{k}^{p} ]] =
\frac{-g_{tt}(R)}{R^{2}} \, 
[(\hat{L}^{p})^{2},\hat{A}_{k}^{p}] \,\, ,
\label{quantum_constraint}
\end{equation}
with $p \in \{ sc,vec \}$ and $k \in \{r, \theta, \phi \}$. The validity of proposed relations in Eq.~\eqref{quantum_constraint} will be justified by their classical limit. As will be seen in the end of this section, from Eq.~\eqref{quantum_constraint} classical field equations will be obtained. From these equations, in the geometrical optics limit, the classical constraint in Eq.~\eqref{classical_constraint} can be recovered. It follows from the quantum constraints in Eqs.~\eqref{quantum_constraint}, considering the results in Eqs.~\eqref{H_a_adagger}--\eqref{L2_a_adagger_vec}, that the scalar and vector mode photon energies $\epsilon$ assume values in discrete sets $\{\epsilon_{\ell}^{p},\ell=1,2,\dots\}$ labeled by $\ell$, where
\begin{eqnarray}
\epsilon_{\ell}^{sc}  & = & \frac{\sqrt{-g_{tt}(R)}}{R} \,
\sqrt{\ell\left( \ell+1\right)  } \,\, ,
\label{quantum_spectrum} \\
\epsilon_{\ell}^{vec}  & = & \frac{\sqrt{-g_{tt}(R)}}{R} \,
\sqrt{\ell\left( \ell+1\right) - 1} \,\, .
\label{quantum_spectrum_vec}
\end{eqnarray}
These are the photon sphere energy spectra of the electromagnetic scalar and vector sectors. The relations~\eqref{quantum_spectrum} and \eqref{quantum_spectrum_vec}, respectively for the scalar and vector mode quanta, are the quantum version of the classical constraint in Eq.~\eqref{classical_constraint}.

\subsection{Geometrical optics limit}

As a consistency check, let us interpret the obtained results in the geometrical optics limit. The results in this section are important for the proper justification of the proposed quantum constraints in Eq.~\eqref{quantum_constraint}.

In the classical limit, the quantum constraints imply that the classical scalar and vector mode potentials $A_{r}^{sc}$, $A_{\theta}^{vec}$, and $A_{\phi}^{vec}$ satisfy Klein-Gordon-type equations (at the photon sphere),
\begin{equation}
\frac{\partial^{2} \Phi}{\partial t^{2}} = 
\frac{-g_{tt}(R)}{R^{2}} \, \tilde{\nabla}^{2} \Phi \,\, ,
\label{classical_field_equation}
\end{equation}
where $\tilde{\nabla}^{2}$ is the Laplace operator on $S^{2}$,
\begin{equation}
\tilde{\nabla}^{2} = 
\frac{1}{\sin^{2}\theta} \, \frac{\partial^{2}}{\partial\phi^{2}} + \frac{1}{\sin\theta}\,\frac{\partial}{\partial\theta}\,\left(\sin\theta\,\frac{\partial}{\partial\theta}\right) \,\, .
\label{nabla_S2}
\end{equation}
Taking the eikonal ansatz,
\begin{equation}
\Phi = \Phi_{0} \, e^{ \textrm{i} \, S } \,\, ,
\end{equation}
with the appropriate eikonal conditions \cite{wald1984general}, we obtain
\begin{equation}
\nabla_{\mu} S \, \nabla^{\mu} S = 0 \,\, .
\label{eikonal_equation}
\end{equation}
It is straightforward to show that not only is $\nabla_{\mu}S$ a null vector, as indicated in the eikonal equation \eqref{eikonal_equation}, but also that its integral curves are null geodesics. 

Moreover, in the eikonal limit we have $\ell \gg 1$, and therefore the scalar and vector electromagnetic spectra coincide in this limit,
\begin{equation}
\left( \epsilon_{\ell} \right)^{2} = -g_{tt}(R) \frac{\ell^{2}}{R^{2}} \,\, ,
\label{quantum_spectrum_eikonal}
\end{equation}
as seen from Eqs.~\eqref{quantum_spectrum} and \eqref{quantum_spectrum_vec}. Relation~\eqref{quantum_spectrum_eikonal} shows that in the eikonal limit, a (scalar or vector) photon in a light ray, with orbital angular momentum $\ell$ and energy $\epsilon_{\ell}$, obeys the classical constraint in Eq.~\eqref{classical_constraint}, as it should by consistency with the geometrical optics limit.

\section{Thermodynamics of photon spheres}
\label{sec_thermodynamics}

In the treatment for the photons introduced in Secs.~\ref{sec_lightspheres} and \ref{sec_spectrum}, the quantum electromagnetic field was assumed to be free, allowing for no direct coupling among the photons in the photon sphere. Still, in many astrophysical situations of interest, some interchange of photons from the photon sphere with the surrounding environment is expected. In fact, perturbations in the photon sphere would take away photons from this region, which is consistent with a finite average lifetime for photons in the light sphere. At the same time, the photon sphere continuously captures external photons, as long as they satisfy the constraint in Eq.~\eqref{classical_constraint}. The astrophysical scenario assumed is a densely populated photon sphere, in dynamical thermal equilibrium with its environment.

We consider then a light sphere populated by photons with a well-defined energy, subjected to the Bose-Einstein statistics, in thermal equilibrium with its surroundings. Moreover, the number of photons is not conserved. Therefore, the total macroscopic energy $U$ of the photon sphere at temperature $T$ is
\begin{equation}
U =
\sum_{p\in \{sc,vec\}} \,\,  \sum_{\ell=1}^{\infty} \,\, \sum_{m=-\ell}^{+\ell}
\frac{\epsilon_{\ell}^{p}}{\exp \left( \frac{\epsilon_{\ell}^{p}}{T} \right) - 1} \,\, .
\end{equation}
Because of spatial isotropy, the sums in $m$ can be readily calculated, 
\begin{equation}
U = \sum_{\ell=1}^{\infty} \,\, 
\left( 2\ell + 1 \right) \, 
\left[
\frac{ \epsilon_{\ell}^{sc}}{ \exp \left( \frac{\epsilon_{\ell}^{sc}}{T}\right) - 1}
+
\frac{ \epsilon_{\ell}^{vec}}{ \exp \left( \frac{\epsilon_{\ell}^{vec}}{T}\right) - 1}
\right] \,\, ,
\label{U_nl}
\end{equation}
where, for convenience, we have separated the contributions of the two modes, presented in Eqs.~\eqref{quantum_spectrum} and \eqref{quantum_spectrum_vec}. 

The relevant macroscopic thermodynamic quantities are defined only if a proper thermodynamic limit can be established \cite{Kuzemsky:2014jna}. In this limit, it is considered the behavior of the photon sphere as its area $\mathcal{A}$ and the number of photons $N$ tend to infinity. It is also required that the density ratio $N / \mathcal{A}$ is bounded. In order to characterize the thermodynamics of the photon sphere, we consider the energy differences 
$\Delta \epsilon^{sc} \equiv \epsilon_{\ell+1}^{sc} - \epsilon_{\ell}^{sc}$ and 
$\Delta \epsilon^{vec} \equiv \epsilon_{\ell+1}^{vec} - \epsilon_{\ell}^{vec}$
of the scalar and vector modes, respectively. In the thermodynamic limit they coincide,
\begin{equation}
\Delta \epsilon =  \Delta \epsilon^{sc} = \Delta \epsilon^{vec}
= \frac{-g_{tt}(R) \, (2\ell + 1)}{2 \epsilon R^{2}}  \,\, ,
\label{Deltaepsilon}
\end{equation}
where it was used in Eq.~\eqref{Deltaepsilon} the fact that $R$ is large and $\Delta\ell/R$ is small in the thermodynamic limit. Therefore, we can rewrite Eq.~\eqref{U_nl} as a Riemann sum, in the form
\begin{equation}
U = \sum_{\epsilon} \,\,
\frac{4 R^{2}}{-g_{tt}(R)} \, \frac{\epsilon^{2}}{\exp\left(\frac{\epsilon}{T}\right)-1} \, \Delta \epsilon \,\, ,
\label{U_l}
\end{equation}
with $\epsilon \in \{ \epsilon_{\ell}^{sc}\} \cup \{ \epsilon_{\ell}^{vec}\}$, according to Eqs.~\eqref{quantum_spectrum} and \eqref{quantum_spectrum_vec}. In terms of the photon sphere area $\mathcal{A} = 4\pi R^{2}$, Eq.~\eqref{U_l} is written as
\begin{equation}
\frac{U}{\mathcal{A}} =
\sum_{\epsilon} \,\, \frac{1}{ -g_{tt}(R) \pi} \, 
\frac{\epsilon^{2}}{\exp \left(\frac{\epsilon}{T}\right)-1} \Delta \epsilon \,\, .
\label{Ull}
\end{equation}

We now take the thermodynamic limit \cite{Kuzemsky:2014jna} of Eq.~\eqref{Ull}, in which $R$ is large and $U/\mathcal{A}$ is bounded. From Eqs.~\eqref{quantum_spectrum} and \eqref{quantum_spectrum_vec}, we see that the maximum value of $\Delta \epsilon$ for a given $\ell$, 
$\Delta \epsilon_{max} = \max \{  \Delta \epsilon^{sc} \} \cup \{  \Delta \epsilon^{vec} \}$, 
can be made arbitrarily small as $R$ is larger. Moreover, the partial sums in Eq.~\eqref{Ull} are well-defined as $\Delta \epsilon_{max}$ is taken arbitrarily smaller. Therefore, the Riemann sum in Eq.~\eqref{Ull} tends to the Riemann integral in the thermodynamic limit (with a fixed value for $\epsilon$). Finally, in the limit of large $\epsilon$, the proper Riemann integral tends to an integral in the form
\begin{equation}
\frac{U}{\mathcal{A}} = \int_{0}^{\infty}\rho(\epsilon)\,d\epsilon \,\, ,
\end{equation} 
with $\rho$ given by
\begin{equation}
\rho(\epsilon) = 
\frac{1}{ -g_{tt}(R) \pi} \frac{\epsilon^{2}}{\exp\left(\frac{\epsilon}{T}\right)-1} 
\,\, .
\label{rho-S2}
\end{equation}
The spectral energy density $\rho$ for the radiation emitted by the photon sphere is qualitatively different from the usual Planck distribution that might be expected, and could provide observational signatures of the photon sphere. This is one of the main results in this paper.

As a side remark, we point out that the spectral distribution in Eq.~\eqref{rho-S2}, although having the same form of the analogous result in the two-torus \cite{Al-Jaber:2003}, was obtained here considering a thermodynamic treatment on a two-sphere, a topological and geometrical distinct object. Moreover, in the two-torus setup, the question of the proper decomposition of the electromagnetic field in the spherical representation is absent. In the two-sphere, this is a nontrivial issue.

From the result in Eq.~\eqref{rho-S2}, we observe that the radiation emitted by a photon sphere should have a distinct profile, when compared to the emission of a usual star. For instance, the emitted total energy of the photon sphere is given by
\begin{equation}
U = \sigma T^{3} \,\, ,
\end{equation}
where $\sigma$ is a constant. This is the modified Stefan-Boltzmann law for the quantum photon sphere.

\section{Final remarks}
\label{sec_final}

In the present work we considered spherically symmetric and asymptotically flat geometries, modeling ultracompact bodies capable of maintaining light in closed orbits. We derived quantum and thermodynamic properties of photon spheres in thermal equilibrium with the environment. The electromagnetic field in the photon sphere was considered in a second quantization scheme, and its energy spectra derived. The associated thermodynamics suggests an observational signature that could be used to distinguish photon spheres from other astrophysical objects.

The results obtained here are very general within the specified premises. For instance, the Einstein field equations are not used, and in this sense only kinematic aspects of gravity are assumed in the present work. A relevant point is that gravity manifests itself only through the redshift factor $-g_{tt}(R)$, according to Eqs.~\eqref{quantum_spectrum}, \eqref{quantum_spectrum_vec}, and \eqref{rho-S2}.

It should be pointed out that the quantum constraint in Eq.~\eqref{quantum_constraint} is not unique. Other (nonequivalent) quantizations, and consequently other dynamics, are possible. Nonetheless the proposal in Eq.~\eqref{quantum_constraint} is ``robust'' in the sense that it correctly gives expected limits, considering both the one-particle sector and the eikonal regime.
In fact, the nonunique character of the quantization procedure is not a peculiarity of our work, but a general issue in the quantization process with Fock spaces. Usually the problem of the vacuum choice, when treating quantum theories in Minkowski spacetime, can be resolved by the imposition of invariance by the action of the Poincar\'e group (and the mass-shell condition). In our case, this procedure can be applied since we have an asymptotic flat limit. Therefore, one argument showing that our constraint is satisfactory is the wave equation \eqref{classical_field_equation}, which corresponds to the field equation for the electromagnetic field in the Feynman gauge. 
This equation is covariant (for each potential component), and tends to a Klein-Gordon equation with the Minkowski metric considering large values of $R$. This suggests that the vacuum of the quantum field theory obtained with Eq.~\eqref{quantum_constraint} is well-defined. Any other quantization, which respects covariance and the mass-shell condition (with a redshift correction), will be unitarily equivalent.

One assumption made in this work is the consideration that the compact objects maintaining the photon sphere have no electric charge. This condition was implicitly used when we postulated that the electromagnetic perturbations do not couple with the gravitational ones \cite{PhysRevD.80.029906}. Still, the treatment of photon spheres around charged compact objects should be possible if the electromagnetic-gravitational compound modes in \cite{Chandra1998mathematical,PhysRevD.41.403,PhysRevD.69.104013} are considered. 

Further generalizations could be made with the consideration of asymptotically de Sitter or anti--de Sitter spacetimes. These could be interesting in cosmological setups or AdS/CFT applications. In fact, the assumption of asymptotically flatness is not needed for any of the classical results presented, and could possibly be relaxed if a proper quantization scheme is used. In this case, the energy definition and the proper choice of observers would become a relevant issue. Work along those lines is currently under way.

\vspace{0.5cm}

\begin{acknowledgments}
This work was partially supported by Conselho Nacional de Desenvolvimento Cient\'{\i}fico e Tecnol\'ogico (CNPq) and Funda\c{c}\~{a}o de Amparo \`{a} Pesquisa do Estado de S\~{a}o Paulo (FAPESP), Brazil.
\end{acknowledgments}


%

\end{document}